# Controlling chaotic itinerancy in laser dynamics for reinforcement learning

**Authors**

Ryugo Iwami,[1]* Takatomo Mihana,[1] Kazutaka Kanno,[1] Satoshi Sunada,[2,3] Makoto Naruse,[4] and Atsushi Uchida[1]*

**Affiliations**

[1]Department of Information and Computer Sciences, Saitama University, 255 Shimo-okubo, Sakura-ku, Saitama City, Saitama 338-8570, Japan.

[2]Faculty of Mechanical Engineering, Institute of Science and Engineering, Kanazawa University, Kakuma-machi Kanazawa, Ishikawa 920-1192, Japan.

[3]Japan Science and Technology Agency (JST), PRESTO, 4-1-8 Honcho, Kawaguchi, Saitama 332-0012, Japan.

[4]Department of Information Physics and Computing, Graduate School of Information Science and Technology, The University of Tokyo, 7-3-1 Hongo, Bunkyo-ku, Tokyo 113-8656, Japan.

*Corresponding author. Emails: r.iwami.692@ms.saitama-u.ac.jp (R. I.); auchida@mail.saitama-u.ac.jp (A. U.)

**Abstract**

Photonic artificial intelligence has attracted considerable interest in accelerating machine learning; however, the unique optical properties have not been fully utilized for achieving higher-order functionalities. Chaotic itinerancy, with its spontaneous transient dynamics among multiple quasi-attractors, can be employed to realize brain-like functionalities. In this paper, we propose a method for controlling the chaotic itinerancy in a multi-mode semiconductor laser to solve a machine learning task, known as the multi-armed bandit problem, which is fundamental to reinforcement learning. The proposed method utilizes ultrafast chaotic itinerant motion in mode competition dynamics controlled via optical injection. We found that the exploration mechanism is completely different from a conventional searching algorithm and is highly scalable, outperforming the conventional approaches for large-scale bandit problems. This study paves the way to utilize chaotic itinerancy for effectively solving complex machine learning tasks as photonic hardware accelerators.

# Introduction

Photonic accelerators provide fast and efficient information processing by using photonic technologies to overcome the limitations of integrated circuit density in semiconductor technologies, that is, the end of Moore's law (1, 2). Notably, photonic accelerators have been discussed for certain dedicated information processing in machine learning, such as deep neural network, and they are expected to contribute to artificial intelligence. Photonic accelerators can be considered as preprocessors that use optical signals combined with electronic computing (1). Recently, conceptually novel principles and technologies of photonic accelerators have been proposed and demonstrated, such as artificial photonic neural networks (3), coherent Ising machines (4), optical pass gate logic (5), photonic reservoir computing (6–8), and photonic decision making (9–13).

In photonic decision making, the multi-armed bandit problem (14) has been examined; the objective of this problem is to maximize the total reward from multiple choices. The multi-armed bandit problem is fundamental to reinforcement learning, where an agent chooses an appropriate action among multiple choices in dynamically changing environments (15,16). For example, the agent selects multiple slot machines with unknown hit probabilities to maximize the total reward. It is important to balance two different actions to solve the multi-armed bandit problem: exploration and exploitation. In exploration, the agent plays the slot machines one by one to search for the slot machine that offers the highest reward, whereas, in exploitation, the agent only plays the slot machine that offers the highest reward, which is estimated from exploration (15). There is a trade-off between exploration and exploitation, known as the exploration–exploitation dilemma, to maximize the total reward. More exploration and less exploitation result in correct estimation of the best slot machine, although the total reward may be reduced. On the contrary, less exploration and more exploitation increase the reward from one of the slot machines estimated as the best slot machine; however, this estimation may be incorrect. The selection of the slot machine with the highest hit probability has been successfully achieved using photonic dynamical systems (9–13). The versatile physical properties of optical processes have been utilized for photonic decision making, chaotically oscillating temporal waveforms of chaotic lasers (9), spontaneous mode switching in a ring-cavity semiconductor laser (11), and lag synchronization of chaos in mutually coupled semiconductor lasers (12).

The scalability of decision making, that is, how to deal with the increasing number of slot machines or choices, is crucial. Reportedly, hierarchical architecture (10) and laser network dynamics (13) accommodate a large number of slot machines. However, these methods have several limitations; for example, the performance depends on the arrangement of the slot machines in the hierarchical architecture (10), and the scalability is less efficient in laser networks (13). Indeed, the scaling exponent, which is a metric used to quantify the scalability of decision making, is limited to a number between 1 and 2 (10, 13), which implies the importance of more efficient decision-making principles to accommodate a large number of choices. In addition, the Lotka–Volterra competition mechanism, which is applicable to ecological systems, has been utilized for decision making (17). Biological species in an ecosystem grow their population by competing for resources, which may be limited, and this interspecific competition mechanism can be modeled using simple ordinary differential equations, namely, the competitive Lotka–Volterra equations. A logarithmic scaling with respect to the number of slot machines has been demonstrated by using this mechanism for decision making. In other words, this mechanism outperforms the software algorithms reported in the literature, as well as the above-mentioned photonic approaches. However, it should be noted that a large number of plays must be assumed to satisfy the analysis condition, indicating that the decision-making method based on this mechanism suffers from practical difficulties.

Chaotic itinerancy has been reported in many interdisciplinary research fields, such as physiological activity (18–20), recurrent neural networks (21–23), neurorobotics (24–26), coupled map lattice (27), and optical turbulence (28). Chaotic itinerancy is a phenomenon where multiple unstable attractors, called quasi-attractors, coexist, and the variables of dynamical systems move around these quasi-attractors (20,21). Chaotic itinerancy is considered essential to understand the emergence of spontaneous activities in the brain (21, 27). In addition, chaotic itinerancy has been used to implement associative memory (29). Recently, spontaneous behavioral switching has been designed by utilizing chaotic itinerancy (30). The implementation of chaotic itinerancy through practical engineering platforms for machine learning is a promising and exciting approach for realizing high functionalities in the brain, such as spontaneous creativity and associative memory.

Chaotic itinerancy has been observed in photonic systems (31, 32), as the chaotic mode competition dynamics among multiple longitudinal modes in a multi-mode semiconductor laser. A variety of nonlinear dynamics and chaotic behaviors in semiconductor lasers with optical feedback and injection have been reported (33–36). Chaotic antiphase dynamics have been experimentally reported in a multi-mode semiconductor laser with optical feedback (37), where the temporal waveforms of the multi-mode semiconductor laser intensities are anti-correlated. In the chaotic mode competition dynamics, the mode with the maximum intensity (called the dominant mode) is dynamically changed in a multi-mode semiconductor laser with optical feedback (38). In addition, the chaotic mode competition dynamics can be strongly influenced by external optical injection (39–41). Thus, the chaotic mode competition dynamics in a multi-mode semiconductor laser could be a suitable tool for achieving an effective spontaneous searching ability for discovering an optimal choice in multiple uncertainties.

In this paper, we propose photonic decision making through chaotic itinerancy control, that is, mode competition dynamics, in a multi-mode semiconductor laser with optical feedback and injection. We solve the multi-armed bandit problem, which is the foundation of reinforcement learning, by utilizing chaotic itinerancy for efficient exploration over many choices. We examine the scalability of the number of choices and demonstrate that the chaotic itinerancy-based method outperforms the UCB1-tuned method, which is one of the most well-known software algorithms. The present study is conducted to investigate chaotic itinerancy in order to utilize the unique physical characteristics of laser dynamics, as well as to address the scalability issues of photonic decision-making principles. To the best of our knowledge, this is the first demonstration of utilizing chaotic itinerancy for accelerating reinforcement learning tasks and establishing a concrete photonic hardware architecture comprising technologically feasible device elements.

## Results
### Multi-mode semiconductor laser with optical feedback and injection

Figure 1 schematically shows the system architecture and dynamics of a multi-mode semiconductor laser with optical feedback and injection. Five longitudinal modes of the multi-mode semiconductor laser are assumed to be excited, whose optical frequencies are denoted as $v_m$ for the $m$-th modal intensity ($m = 1, 2, …, 5$, $v_i < v_j$ for $i < j$). In addition, a single-mode semiconductor laser with an optical frequency $f_m$ is used for the optical injection. The optical output of the single-mode laser is injected into the $m$-th modal intensity with a frequency $v_m$ in the multi-mode semiconductor laser to control the mode competition dynamics, as shown in Fig. 1. $f_m$ is slightly detuned from $v_m$ to achieve injection locking.

We used a numerical model for a multi-longitudinal-mode semiconductor laser with optical feedback (35, 38, 42). This model equation is an extension of the Lang–Kobayashi equations (35, 36, 43), which are well-known numerical model equations for a semiconductor laser with optical feedback. We also added an optical injection term from a single-mode semiconductor laser (44). Modes 1, 2, …, $M$ are assigned from the lower to higher frequency modes. Details of the numerical model used in this study are provided in the Materials and Methods section.

Figure 2 summarizes the temporal waveforms of the optical intensity, $I_m(t) = |E_m(t)|^2$, obtained via numerical simulation. Figure 2(A) shows the temporal waveforms of the modal intensities without optical injection. All the modal intensities oscillate chaotically, and the temporal dynamics of the modal intensities are different. Figure 2(B) shows the temporal waveform of the total intensity, which is the sum of the modal intensities $I_{total}(t) = \Sigma I_m(t)$; the total intensity oscillates chaotically. Figure 2(C) shows the temporal waveforms of the modal intensities when the optical signal is injected into mode 3 with an injection strength of $\kappa_{inj,3} = 6.0$ ns$^{-1}$. The oscillation of mode 3, denoted by the red curve, is enhanced by the optical injection, and the duration of the dominant mode for mode 3 is longer than that for the other modes. Therefore, the oscillations of the other modes are suppressed. Figure 2(D) shows the temporal waveform of the total intensity obtained from Fig. 2(C), and this temporal waveform also exhibits chaotic oscillation. Notably, the optical injection does not result in any significant change in the mean of the total intensity, because the oscillation of mode 3 is enhanced, and those of the other modes are suppressed spontaneously owing to the power conservation in the total intensity (i.e., antiphase dynamics (37)).

Figure 3(A) shows the temporal waveforms of the optical frequency detuning for the total intensity and the five modal intensities without optical injection. The optical frequency detuning of the total and modal intensities from the central frequency is calculated from the phase equations. The optical frequency detuning of the total intensity (black curve) dynamically moves among the different modes, and the transition of the modes occurs spontaneously. Here, we clearly observe chaotic itinerancy among the five modes that are unstable quasi-attractors. The total intensity stays at one of the modes and shifts to the neighboring modes with time. In fact, the optical frequency detuning of the total intensity fluctuates very rapidly, corresponding to the longitudinal mode interval $\Delta \nu = 35.5$ GHz. Figure 3(B) shows the temporal waveforms of the optical frequency detuning for the total intensity and the five modal intensities with optical injection into mode 3 ($\kappa_{inj,3} = 6.0$ ns$^{-1}$). Although chaotic itinerancy is still observed, the total mode remains at mode 3 for a relatively long duration. The residence time of the total intensity in other modes is reduced as shown in Fig. 3(B), compared with that shown in Fig. 3(A), owing to the optical injection to mode 3. In this manner, the chaotic itinerancy can be controlled via optical injection.

We investigated the residence time of the total intensity on one of the modes when chaotic itinerancy occurred without optical injection. The residence time is defined as the duration of the optical frequency of the total intensity (obtained from the phase of the total electric-field amplitude) staying in one mode over one oscillation period, which is 28 ps, determined by a longitudinal mode interval of 35.5 GHz. We measured the residence time for each modal intensity and created a histogram of the residence time to evaluate the ratio of observing certain residence times, followed by evaluating the probability of residence time. Figure 3(C) shows the probability of the residence time of the total intensity for each modal intensity without optical injection; these probabilities are calculated from 10-ms-long temporal waveforms. The curves exhibit a linear reduction as a function of residence time on a semi-logarithmic scale. Therefore, we found an exponential relationship of the residence-time probability as $P = A e^{\beta t}$, where $t$ denotes the residence time, and $A$ and $\beta$ are real numbers. The exponents of the exponential decay $\beta$ are different for the five modes: $\beta = -2.7, -2.0, -1.6, -2.0,$ and $-2.7$ for modes 1, 2, 3, 4, and 5, respectively. The mean residence times $\tau_r = 1/|\beta|$ are 0.37, 0.50, 0.63, 0.50, and 0.37 ns for modes 1, 2, 3, 4, and 5, respectively. $\tau_r$ is the maximum at the central mode (i.e., mode 3), whereas it decreases as the mode becomes closer to the mode at the edge (modes 1 and 5). Therefore, laser dynamics are highly likely to provide a relatively stable residence when the mode is located at the center, whereas it explores other modes when the mode is located far from the central mode.

Figure 3(D) shows the probability of the residence time of the total intensity for each modal intensity under optical injection in mode 3. The residence time in mode 3 is enhanced via optical injection, and the absolute value of the slope of the probability curve decreases (indicated by the red curve). In contrast, the residence time in other modes is reduced, and the absolute value of the slope is increased. It is worth noting that different slopes are observed in the regions of short (< 1 ns) and long (> 1 ns) residence times for all the modes. Therefore, the statistical characteristics of the chaotic itinerancy can be altered via optical injection.

We measured the change in the dominant mode ratio (38) under optical injection. The dominant mode ratio represents the ratio of the probability at which the $m$-th modal intensity reaches the maximum value over time. The dominant mode ratio $DMR_m$ of mode $m$ is defined as follows (38),

$$DMR_m = \frac{1}{d}\sum_{l=1}^{d} D_m(l) \qquad (1)$$

where $d$ is the total number of sampling points and corresponds to the length of the temporal waveform used for the calculation. $D_m(l)$ is the function that returns 1 if mode $m$ has the maximum value among the other modes at the $l$-th sampling point, and 0 otherwise.

Figure 4(A) shows the dominant mode ratio as a function of the optical injection strength subjected to mode 3, ranging from 0.0 ns$^{-1}$ to 15.0 ns$^{-1}$. The dominant mode ratio is calculated from the temporal waveforms over 20 μs for each optical injection strength. In Fig. 4(A), the dominant mode ratio of mode 3 (red line) drastically increases when the optical injection strength becomes greater than approximately 2.5 ns$^{-1}$. When the optical injection strength is greater than approximately 8.0 ns$^{-1}$, the dominant mode ratio of mode 3 is one, meaning that mode 3 becomes the perfectly dominant mode. Figure 4(B) shows the dominant mode ratio as a function of the optical injection strength subjected to mode 1 (outermost mode). It is worth noting that the optical frequency detuning between mode 1 and the optical injection signal must be adjusted carefully to match the optical feedback phase (see the Materials and Methods section). Under this condition, the enhancement of the mode excitation (i.e., dominant mode control based on optical injection) is effective even for the side modes with a smaller gain. From such characterizations, the probability of a certain mode being the dominant mode can be configured by changing the optical injection strength. In other words, we found that the mode competition dynamics can be controlled by engineering the optical injection into particular modes.

**Decision making using multi-mode semiconductor laser with optical feedback and injection**

Here, we describe the decision-making principle for solving the multi-armed bandit problem by using the aforementioned multi-mode semiconductor laser dynamics. Figure 5 shows a schematic diagram for photonic decision making using a multi-mode semiconductor laser with optical feedback and injection. We solve the multi-armed bandit problem with $M$ slot machines by assigning each modal intensity of $M$ longitudinal modes of the multi-mode semiconductor laser to each slot machine, as shown in Fig. 5. The dominant mode is determined by comparing the modal intensities at a sampling interval, and the slot machine corresponding to the dominant mode is selected. After the slot machine is selected, the optical injection strength is controlled using the tug-of-war method (45–48), according to the result of the slot machine selection. We assume that the result of the slot machine selection is "hit" or "miss" with the hit probability $P_m$ (miss probability: $1 - P_m$) for the slot machine $m$. The rewards of hit and miss correspond to 1 and 0, respectively (binary rewards; see the Materials and Methods section for details of the decision-making algorithm).

We performed decision making by controlling the optical injection strength of single-mode semiconductor lasers. The mode corresponding to a large evaluation value $X_m(p)$ (see the Materials and Methods section) can be enhanced via optical injection, because the slot machine with a large $X_m(p)$ is evaluated as a good slot machine. The optical injection strength for mode $m$ is controlled at the $p$-th play, as follows:

$$\kappa_{inj,m} = \begin{cases} \kappa_{inj,max} & (kX_m(p) > \kappa_{inj,max}) \\ kX_m(p) & (0 \leq kX_m(p) \leq \kappa_{inj,max}) \\ 0 & (kX_m(p) < 0) \end{cases} \quad (2)$$

where $k$ is the coefficient that adjusts the optical injection strength, and $\kappa_{inj,max}$ is the upper limit of the optical injection strength. In this study, we set $k = 0.1$ ns$^{-1}$ and $\kappa_{inj,max} = 15.0$ ns$^{-1}$.

We examined the decision-making performance when the number of slot machines $M$ was changed. In this case, the modal intensities are sampled at a sampling interval of 0.1 ns and subsequently compared. Thus, the slot machine is selected every 0.1 ns. The hit probability of slot machine 1 is set to 0.9 and those of the other slot machines are set to 0.7. The slot machine $i$ is assigned to mode $i$ in the multi-mode laser with $M$ modes. Therefore, the slot machine with the maximum hit probability is assigned to the outermost mode (mode 1). Figure 6(A) shows the results of slot machine selection as the number of plays is increased for $M = 5$. All the slot machines are selected almost randomly when the number of plays is small. Slot machine 1 is selected more frequently as the number of plays increases. After approximately the 800-th play, only slot machine 1 is selected. Therefore, an accurate slot-machine selection is accomplished. Figure 6(B) depicts the case in which the number of slot machines is large, $M = 129$. Evidently, in the initial phase, the slot machines corresponding to the modes near the central frequency are extensively selected. The selection range is then expanded to other modes as the number of plays increases. Finally, slot machine 1, which is the highest hit probability machine, can be selected successfully after approximately the 6000-th play.

Further, we investigated the statistical characteristics of decision-making performance. As shown in Fig. 6(A), consecutive 2000 plays are conducted and repeated every 500 cycles to evaluate a statistical measure of the correct decision rate (CDR) (9–13), which is defined as follows:

$$CDR(p) = \frac{1}{S}\sum_{l=1}^{S} C(l,p) \tag{3}$$

where $S$ is the total number of cycles; $C(l, p)$ is a function that returns 1 if the slot machine with the highest hit probability is selected at the $p$-th play and the $l$-th cycle, and 0 otherwise; and $CDR(p)$ represents the average rate in $S$ cycles, i.e., the rate at which the slot machine with the highest hit probability is selected at the $p$-th play.

Figure 7(A) shows the CDR as the number of plays is changed when the number of slot machines $M$ ranges from 3 to 513. More precisely, $M$ is specified by $2^n + 1$, where $n$ is a positive integer ranging from 1 to 9. The hit probability of slot machine 1 is set to 0.9 and those of the other slot machines are set to 0.7 for all $M$. The correct decision exhibits a smaller value when the number of plays is small and gradually increases as the number of plays increases. Indeed, the CDR approaches one as the number of plays increases for all $M$, as observed in Fig. 7(A), implying that the highest-hit-probability slot machine is accurately selected. In addition, the curves shown in Fig. 7(A) are almost equidistant on the horizontal logarithmic scale as the number of slot machines $M$ increases in the form of $2^n + 1$. This result indicates that the scaling law between the number of plays for correct decisions and the number of slot machines can be obtained from these curves.

Furthermore, we investigated a well-known quantitative measure of regret to evaluate the decision-making performance. Regret at the $p$-th play is defined as the difference between the ideal (maximum) total reward and the actual reward obtained by the plays, as follows (49, 50):

$$Regret(p) = p\, P_{max} - \frac{1}{S}\sum_{l=1}^{S}\sum_{m=1}^{M} \left(P_m\, S_{l,m}(p)\right) \tag{4}$$

where $P_{max}$ is the maximum hit probability, $S$ is the total number of cycles, $P_m$ is the hit probability of slot machine $m$, and $S_{l,m}(p)$ is the number of selections for slot machine $m$ at the $l$-th cycle until the $p$-th play. A smaller regret indicates a better decision-making performance. In this study, we set $S = 500$ to evaluate the regret. Figure 7(B) shows the regret as the number of plays changes when the number of slot machines $M$ ranges from 3 to 513. The regret curves increase with an increase in the plays and saturate at certain values for all $M$. The saturation of the curves indicates that correct decision making is achieved after a sufficient number of plays are conducted. It is also worth noting that the curves are equidistantly distributed on a logarithmic scale; hence, a scaling law can be obtained from these curves.

**Scalability of decision-making performance**

We investigated the scalability of the decision-making performance when the number of slot machines is changed. We analyzed the number of plays when the CDR reaches 0.95, as shown in Fig. 7(A), to examine scalability (10, 13). The red curve in Fig. 8(A) shows the scalability of the number of plays required for the CDR of 0.95 (denoted as $N_{play}$) as the number of slot machines $M$ is changed. $N_{play}$ is proportional to $M$ on a logarithmic scale for both the vertical and horizontal axes. Thus, we approximate the curve shown in Fig. 8(A) by using the power law: $N_{play} = a \cdot M^{\gamma}$. We obtain $N_{play} = 318\, M^{0.70}$, that is, the scaling exponent $\gamma$ is 0.70, which is less than 1. Therefore, our decision-making method using the multi-mode semiconductor laser is suitable for solving the multi-armed bandit problem with a large number of slot machines, whereas the scaling exponents for the previous methods are between 1 and 2 (10, 13).

We compared the scalability of our scheme with that of the UCB1-tuned (upper confidence bound 1-tuned) method (49), which is a well-known software algorithm for solving the multi-armed bandit problem. The blue curve shown in Fig. 8(A) indicates the number of plays ($N_{play}$) required to achieve a CDR of 0.95, as the number of slot machines $M$ increases for the UCB1-tuned method. When the number of slot machines is small, UCB1-tuned shows a superior performance because it requires fewer plays for correct decision making. However, when the number of slot machines increases, our method is superior in achieving the correct decision making with fewer plays. We approximate these curves by using the power laws. The curve for the UCB1-tuned method is approximately given by $N_{play} = 93 M^{1.06}$; that is, the scaling exponent of the power law for the UCB1-tuned method is 1.06, which is larger than that for the proposed method of 0.70. The difference in the scaling exponents can significantly influence the search capability of the best choice, particularly for an extremely large-scale multi-armed bandit problem. For example, the proposed method enables us to determine the best choice at a 2.7 times faster rate, compared to the UCB1-tuned method for $M = 500$. The best choice can be determined 6.3 times faster for $M = 5000$.

We also obtained the scaling exponent from the results of regret, as shown in Fig. 7(B). Figure 8(B) shows the regret at the number of plays $p = 60000$ as the number of slot machines $M$ increases for the multi-mode laser (red) and UCB1-tuned method (blue). We approximate the curves shown in Fig. 8(B) by a power law $Regret = b \cdot M^{\Gamma}$ and obtain $Regret = 33.7 M^{0.73}$ for the multi-mode laser and $Regret = 9.28 M^{1.06}$ for the UCB1-tuned method. The scaling exponents are $\Gamma = 0.73$ and 1.06 for the multi-mode laser and UCB1-tuned method, respectively. Therefore, the scaling exponent for the multi-mode laser is smaller than that for the UCB1-tuned method, even when using the regret measure. A better decision-making performance can be achieved using the multi-mode laser for over 100 slot machines, as shown in Fig. 8(B).

To understand the improvement in the performance of our method, we compared the characteristics of the slot-machine selection between the proposed method and the UCB1-tuned method. We calculated the Shannon entropy from the probabilities of the number of selections for each slot machine to evaluate the bias of the selection probabilities (51) (see the Materials and Methods section for details of the calculation of the Shannon entropy). A bias in the slot machine selection exists if the entropy is small. In particular, only one slot machine is always selected if the entropy is 0. Figure 9(A) shows the Shannon entropy of the probabilities of the number of selection for each slot machine when the number of slot machines is $M = 5$, as the number of plays is changed, using the multi-mode laser and UCB1-tuned method. The entropy for the UCB1-tuned method (blue curve in Fig. 9(A)) decreases gradually and almost monotonically as the number of plays increases on a double logarithmic scale. In contrast, the entropy for the multi-mode laser (red curve in Fig. 9(A)) is almost constant when the number of plays is small, and it suddenly decreases after approximately the 600-th play. In other words, the multi-mode laser system explores a variety of selections at the beginning, and then, accurate decision making is suddenly accomplished, as demonstrated in Fig. 6(A). We interpret that the constant entropy corresponds to the exploration procedure, where all the slot machines are selected equally, whereas the sudden dropout of the entropy corresponds to the exploitation procedure, where the slot machine selection is suddenly "accelerated" after sufficient exploration owing to the chaotic mode competition dynamics.

Figure 9(B) shows the entropy when the number of slot machines is increased to $M = 129$. The entropy of the UCB1-tuned method also decreases gradually as the number of plays increases, similar to the case of $M = 5$ in Fig. 9(A). Similarly, the entropy for the multi-mode laser is constant at the beginning; in fact, it slightly increases, which coincides with the expansion of the selection range, as observed in Fig. 6(B). The entropy decreases rapidly after approximately the 5000-th play, realizing a low entropy much faster than that achieved using the UCB1-tuned method. Based on the results shown in Fig. 9(B), we consider that the exploration process is similar for the multi-mode laser system and UCB1-tuned method. However, the exploitation is accelerated by the multi-mode laser method. Therefore, the proposed method using a multi-mode laser outperforms the UCB1-tuned software algorithm. The multi-mode laser can accelerate the process of slot machine selection when a large number of slot machines exist owing to the fast convergence of the chaotic mode competition dynamics.

**Discussion**

Our results show that chaotic itinerancy (i.e., the mode competition dynamics) in a multi-mode semiconductor laser provides efficient exploration and exploitation to identify the slot machine with the largest reward. (i.e., best choice). This partly originates from the chaotic "partition" of lasing energy which is highly sensitive to external stimuli (e.g., optical injection) (35). Our results indicate that optical injection can result in an efficient energy concentration in a particular mode corresponding to the best choice, even when many modes are involved in the lasing state. In other words, one of the multiple modes can be enhanced easily via optical injection, whereas the other modes are suppressed, because the oscillation of each modal intensity is very weak in the presence of a large number of modes. The convergence of the total intensity to one of the modes, that is, a quick transition from the exploration phase to the exploitation phase, can be accelerated even for a low optical-injection energy, which suggests the high feasibility of the implementation in real hardware.

Our method using multi-mode laser dynamics outperforms the UCB1-tuned algorithm when the number of slot machines is very large (over 100). The UCB1-tuned algorithm selects slot machines in parallel based on the confidence bound, which gradually decreases the entropy; however, acceleration cannot be induced. Therefore, our method based on multi-mode laser dynamics can select the correct slot machine much faster than the UCB1-tuned algorithm when the number of slot machines is large. Indeed, the scaling exponent of the proposed chaotic itinerancy-based method is 0.70, indicating the advantage of the proposed method under a large number of slot machines, compared to the existing software algorithms and other photonic methods with a scaling exponent of approximately one or more (e.g., the exponent is 1.06 in the UCB1-tuned algorithm, whereas the exponents are 1.16 and 1.85 in the photonic methods reported in (10) and (13), respectively). The identification of the best choice from many choices with unknown rewards is crucial in practical applications such as continuous problems, including online auction or routing (52) or channel searching in wireless and optical communications (53, 54). The proposed photonic approach may open a novel pathway for solving such large-scale bandit problems.

This scheme utilizes the combination of a physical competition mechanism in multi-mode laser dynamics and software control based on the tug-of-war method. Thus, improving the software-control algorithm could improve the decision-making performance, including the hit probability estimation (see the Materials and Method section). In addition, the mode competition dynamics could be tuned by changing the laser parameters, such as the injection current, feedback strength, and feedback delay time, to optimize the decision-making performance. Notably, the feasibility of the experimental implementation of this scheme is another issue. In the experiment, the number of single-mode lasers required for optical injection would increase as the numbers of modes and slot machines are increased. This scaling issue can be solved by using a multi-mode laser, optical wavelength filters, and optical amplifiers, instead of using a large number of single-mode lasers. Further, the optical injection components could also be implemented in a photonic integrated circuit for miniaturizing the optical injection control system. The experimental implementation of this scheme would be our future work.

Our decision-making method could be applied to other nonlinear dynamical systems that produce chaotic itinerancy. The spontaneous searching ability supported by chaotic itinerancy is extremely promising for solving complex machine-learning tasks, as well as for understanding the spontaneous activities of the brain. The engineering design of chaotic itinerancy is crucial to maximize the searching ability using control techniques (30). The combination of chaotic itinerancy and control techniques results in a new research direction for machine-learning applications.

In this study, we demonstrated photonic decision making for solving multi-armed bandit problems by controlling the chaotic itinerancy, known as the mode competition dynamics, in a multi-longitudinal-mode semiconductor laser with optical feedback and injection. We confirmed that the mode competition dynamics in a multi-mode semiconductor laser can be controlled via optical injection in a specific mode. We assigned one slot machine to each modal intensity and controlled mode competition dynamics, based on the results of the slot machine selection. We solved the multi-armed bandit problem with up to 513 slot machines and evaluated the decision-making performance using the CDR and regret. We found that our decision-making scheme shows excellent scalability between the number of plays for correct decision making and the number of slot machines in the form of a power law with a scaling exponent of 0.70, which is superior to that of the well-known UCB1-tuned software algorithm. We investigated the entropy of selection probabilities and found that our decision-making scheme provides fast and efficient decision making for the correct slot machine when the number of slot machines is larger than 100. Our method using multi-mode laser dynamics can enhance the decision-making performance under a large number of choices.

To conclude, this study demonstrated that chaotic itinerancy in multi-mode laser dynamics is a promising resource for solving machine learning tasks as photonic accelerators. The proposed chaotic itinerancy-based principle exploits the high-bandwidth attributes of light as well as complex laser dynamics, which are manifested by the residence time statistics and entropy analysis. The physical properties and architectural design allow an efficient and accelerated bandit-problem solving. Based on the insights gained through the present study, the proposed method that combines chaotic itinerancy and complex laser dynamics can be extended to solve higher-order problems and complex machine learning tasks in the future.

## Materials and Methods

### Numerical model of multi-mode semiconductor laser

The numerical model of the multi-mode semiconductor laser with $M$ longitudinal modes is described by the following deterministic equations (38, 42):

$$\frac{dE_m(t)}{dt} = \frac{1-i\alpha}{2}\left\{\frac{G_m[N(t)-N_0]}{1+\varepsilon\sum_{u=1}^{M}|E_u(t)|^2} - \frac{1}{\tau_p}\right\}E_m(t) \\ + \kappa E_m(t-\tau)\exp(i\omega_m\tau) + \kappa_{inj,m}A_s\exp(-i2\pi\Delta f_m t) \qquad (5)$$

$$\frac{dN(t)}{dt} = J - \frac{N(t)}{\tau_s} - \sum_{v=1}^{M} \left\{ \frac{G_v[N(t) - N_0]|E_v(t)|^2}{1 + \varepsilon \sum_{u=1}^{M}|E_u(t)|^2} \right\} \quad (6)$$

$$G_m = G_N \left[ 1 - \frac{(\nu_m - \nu_{m_c})^2}{\Delta \nu_g^2} \right] \quad (7)$$

where $E_m(t)$ represents the slowly varying complex electric field amplitude of the $m$-th longitudinal mode, and $N(t)$ represents the carrier density. In Eqs. (5) and (6), $i$ is an imaginary unit; $\alpha$ is the linewidth enhancement factor; $G_m$ is the gain coefficient of the $m$-th mode; $N_0$ is the carrier density at transparency; $\varepsilon$ is the gain saturation coefficient; $\tau_p$ is the lifetime of the photon; $\tau$ is the carrier lifetime; $\kappa$ is the optical feedback strength of the multi-mode semiconductor laser; $\kappa_{inj,m}$ is the optical injection strength from the single-mode semiconductor laser to the $m$-th mode of the multi-mode semiconductor laser; $\tau$ is the round-trip time of light in the external cavity of the multi-mode laser; and $\omega_m$ is the angular frequency of the $m$-th mode ($\omega_m = 2\pi \nu_m$). Here, $A_s$ represents the steady-state solution of the electric-field amplitude of a single-mode semiconductor laser without optical feedback and injection (44). $\Delta f_m$ is the initial optical frequency detuning between the single-mode semiconductor laser with frequency $f_m$ and the $m$-th mode of the multi-mode laser ($\Delta f_m = f_m - \nu_m$), and $J$ is the injection current. The gain coefficient of the $m$-th mode is approximated by a parabolic gain profile, as shown in Eq. (7), where $G_N$ is the gain coefficient at the central mode $m_c = (M + 1) / 2$, where $M$ is assumed to be an odd integer. $\Delta \nu_g$ is the frequency width of the gain profile. We define the frequency of the $m$-th mode as $\nu_m = \nu_{mc} + (m - m_c)\Delta \nu$, where $\nu_{mc}$ is the frequency of the central mode $m_c$, and $\Delta \nu = 35.5$ GHz is the frequency spacing among the longitudinal modes. In this model, no spontaneous emission noise is included to investigate the dynamics of deterministic chaotic itinerancy. The parameter values used in this study are listed in Table 1.

The frequency width of the gain profile is set to $\Delta \nu_g = 141.9$ THz (which corresponds to the wavelength width 1000 nm) to investigate the scaling properties for a large number of modes and slot machines. Here, $\Delta \nu_g = 141.9$ THz may be too wide for the gain width of an actual semiconductor laser (the typical gain width of semiconductor lasers is several tens of THz (or 10–100 nm) (38, 42, 55)). It is necessary to set a large gain width to increase the number of modes (up to 513) that exhibit chaotic mode competition dynamics via optical feedback without the noise term. This is because the modes near the central frequency of the gain curve are raised if the noise term is not included, and the mode coupling is included only via the carrier density (42, 56). We confirmed that similar results described in the main text are obtained when using a smaller gain width, $\Delta \nu_g = 12.70$ THz (wavelength width 100 nm), and a smaller number of modes as realistic parameter values.

**Tuning of frequency detuning of optical injection**
To perform decision making, the initial optical frequency detuning between each mode and the optical injection signal must be adjusted. The difference in the optical feedback phase for different longitudinal modes causes a change in the dominant mode ratio. Moreover, the optical injection strength, required for obtaining a large dominant mode ratio, depends on the optical feedback phase of the injection signal (57). Considering the mode spacing $\Delta \nu$ and the difference in the optical feedback phase among the modes $2\pi\Delta\nu\tau$, we modify $\Delta f_m$ with the phase shift among different modes $\Phi_{adjust,m}$ as follows:

$$\Phi_{adjust,m} = 2\pi(m_c - m)\Delta\nu\tau \quad (\text{mod } 2\pi) \quad (8)$$

$$\Delta f_m = \Delta f_{m_c} + \frac{1}{\tau}\frac{1}{2\pi}\Phi_{adjust,m} \quad (9)$$

where Eq. (8) represents the phase shift to match the optical feedback phase of the $m$-th mode with that of the central mode $m_c$. In Eq. (9), the initial optical frequency detuning can be adjusted to compensate for the phase shift calculated using Eq. (8). We fix the initial frequency detuning for the central mode at $\Delta f_{mc} = -4.0$ GHz, and the $\Delta f_m$ values of the other modes are adjusted using Eqs. (8) and (9) to obtain similar characteristics of optical injection among the different modes, as shown in Figs. 4(A) and 4(B).

**Decision-making algorithm**

We describe the tug-of-war method for $M$-slot machines used in this study (9–12, 45, 46, 48). The evaluation value $X_m(p)$ for the slot machine $m$ at the $p$-th play originates from the displacement of the amoeba branch in the tug-of-war method; the initial value is $X_m(0) = 0$. If the slot machine $s$ is played at the $p$-th play and the result is hit, $X_m(p)$ is changed as follows:

$$X_m(p) = \begin{cases} X_m(p-1) + \Delta(p) & (m = s) \\ X_m(p-1) - \dfrac{\Delta(p)}{M-1} & (m \neq s) \end{cases} \quad (10)$$

where $\Delta(p)$ represents the change in the hit case. If the slot machine $s$ is played at the $p$-th play, and the result is missed, then $X_m(p)$ changes as follows:

$$X_m(p) = \begin{cases} X_m(p-1) - \Omega(p) & (m = s) \\ X_m(p-1) + \dfrac{\Omega(p)}{M-1} & (m \neq s) \end{cases} \quad (11)$$

where $\Omega(p)$ represents the amount of change in the miss case. The amounts of changes $\Delta(p)$ and $\Omega(p)$ differ between hit and miss because either of them needs to be prioritized based on the parameter settings of the hit probabilities (10). The sub-optimal values of $\Delta(p)$ and $\Omega(p)$ are determined by the sum of the two highest hit probability values (47, 48). However, the hit probabilities are unknown initially, and $\Delta(p)$ and $\Omega(p)$ need to be estimated using the results of the slot machine selection as follows (12):

$$\Delta(p) = 2 - \left(\hat{P}_{top1}(p) + \hat{P}_{top2}(p)\right) \quad (12)$$

$$\Omega(p) = \hat{P}_{top1}(p) + \hat{P}_{top2}(p) \quad (13)$$

where $\hat{P}_{top1}(p)$ and $\hat{P}_{top2}(p)$ represent the highest and second-highest estimated hit probabilities, respectively, based on the results of the slot machine selection.

In the previous studies (9–13), the hit probability of the slot machine $s$ was estimated by the ratio between the number of hits and the number of total plays for the slot machine $s$. However, this method may result in incorrect estimation at the beginning of the selection process when the number of slot machines is large because $\Delta(p)$ and $\Omega(p)$ only require the two highest estimated hit probabilities. In addition, $\Delta(p)$ and $\Omega(p)$ are estimated from a limited number of the slot-machine selection results, and incorrect exploitation may be performed if all the slot machines are not selected. To overcome this limitation, we modify the estimation method of the hit probabilities at the $p$-th play for the $m$-th slot machine ($\hat{P}_m(p)$) as follows:

$$\hat{P}_m(p) = \begin{cases} \dfrac{R_m(p)}{S_m(p) + 1} & (S_m(p) \neq 0) \\ P_{unknown}(p) & (S_m(p) = 0) \end{cases} \quad (14)$$

$$P_{unknown}(p) = \hat{P}_{top1}(p) \quad (15)$$

where $R_m(p)$ and $S_m(p)$ represent the number of hits and plays for slot machine $m$ until the $p$-th play, respectively. An extremely high estimation of the hit probabilities can be avoided by adding 1 to the denominator in Eq. (14) when the number of plays is small. In addition, the estimated hit probability for the slot machine that has never been selected until the $p$-th play is set to $P_{unknown}(p) = \hat{P}_{top1}(p)$, where $\hat{P}_{top1}(p)$ is the maximum value of $\hat{P}_m(p)$ in Eq. (14), to facilitate exploration for all the slot machines. According to Eqs. (14) and (15), the second-highest estimated hit probability becomes the same as the highest estimated hit probability until all the slot machines are selected, resulting in the selection of all the slot machines. In general, $P_{unknown}(p)$ can be set to a constant value in advance; however, the decision-making performance depends on this constant value. Our method is useful because $P_{unknown}(p)$ is adaptively changed based on the hit probabilities estimated by the exploration.

**Shannon entropy of probabilities of slot-machine selection**
We calculate the Shannon entropy from the probabilities of slot-machine selection. We calculate the selection probability of the slot machine *m* until the *p*-th play using window size *W* as follows:

$$P_{sel,m}(p) = \frac{1}{W} \sum_{l=p-W+1}^{p} Sel_m(l) \qquad (16)$$

where $Sel_m(l)$ is a function that returns 1 if slot machine *m* is selected at the *l*-th play, and 0 otherwise. The window size *W* was set to 5*M*, where *M* is the number of slot machines. The Shannon entropy of the selection probability at the *p*-th play is calculated as follows:

$$H(p) = -\sum_{l=1}^{M} P_{sel,l}(p) \log_2 \left( P_{sel,l}(p) \right) \qquad (17)$$

We calculated the Shannon entropy for each cycle and averaged it over 500 cycles, as shown in Fig. 9.

**References**
1. K. Kitayama, M. Notomi, M. Naruse, K. Inoue, S. Kawakami, A. Uchida, Novel frontier of photonics for data processing—photonic accelerator, *APL Photonics* **4**, 090901 (2019).
2. B. J. Shastri, A. N. Tait, T. F. de Lima, W. H. P. Pernice, H. Bhaskaran, C. D. Wright, P. R. Prucnal, Photonics for artificial intelligence and neuromorphic computing, *Nat. Photonics* **15**, 102-114 (2021).
3. Y. Shen, N. C. Harris, S. Skirlo, M. Prabhu, T. Baehr-Jones, M. Hochberg, X. Sun, S. Zhao, H. Larochelle, D. Englund, M. Soljačić, Deep learning with coherent nanophotonic circuits, *Nat. Photonics* **11**, 441-446 (2017).
4. T. Inagaki, Y. Haribara, K. Igarashi, T. Sonobe, S. Tamate, T. Honjo, A. Marandi, P. L. McMahon, T. Umeki, K. Enbutsu, O. Tadanaga, H. Takenouchi, K. Aihara, K. Kawarabayashi, K. Inoue, S. Utsunomiya, H. Takesue, A coherent ising machine for 2000-node optimization problems, *Science* **354**, 603-606 (2016).
5. T. Ishihara, A. Shinya, K. Inoue, K. Nozaki, M. Notomi, An integrated nanophotonic parallel adder, *ACM J. Emerg. Technol. Comput. Syst.* **14**, 26 (2018).
6. L. Larger, M. C. Soriano, D. Brunner, L. Appeltant, J. M. Gutierrez, L. Pesquera, C. R. Mirasso, I. Fischer, Photonic information processing beyond turing: an optoelectronic implementation of reservoir computing, *Opt. Express* **20**, 3241-3249 (2012).
7. D. Brunner, M. C. Soriano, C. R. Mirasso, I. Fischer, Parallel photonic information processing at gigabyte per second data rates using transient states, *Nat. Commun.* **4**, 1364 (2013).
8. K. Takano, C. Sugano, M. Inubushi, K. Yoshimura, S. Sunada, K. Kanno, A. Uchida, Compact reservoir computing with a photonic integrated circuit, *Opt. Express* **26**, 29424-29439 (2018).

9. M. Naruse, Y. Terashima, A. Uchida, S.-J. Kim, Ultrafast photonic reinforcement learning based on laser chaos, *Sci. Rep.* **7**, 8772 (2017).
10. M. Naruse, T. Mihana, H. Hori, H. Saigo, K. Okamura, M. Hasegawa, A. Uchida, Scalable photonic reinforcement learning by time-division multiplexing of laser chaos, *Sci. Rep.* **8**, 10890 (2018).
11. R. Homma, S. Kochi, T. Niiyama, T. Mihana, Y. Mitsui, K. Kanno, A. Uchida, M. Naruse, S. Sunada, On-chip photonic decision maker using spontaneous mode switching in a ring laser, *Sci. Rep.* **9**, 9429 (2019).
12. T. Mihana, Y. Mitsui, M. Takabayashi, K. Kanno, S. Sunada, M. Naruse, A. Uchida, Decision making for the multi-armed bandit problem using lag synchronization of chaos in mutually coupled semiconductor lasers, *Opt. Express* **27**, 26989-27008 (2019).
13. T. Mihana, K. Fujii, K. Kanno, M. Naruse, A. Uchida, Laser network decision making by lag synchronization of chaos in a ring configuration, *Opt. Express* **28**, 40112-40130 (2020).
14. H. Robbins, Some aspects of the sequential design of experiments, *Bull. Am. Math. Soc.* **58**, 527-535 (1952).
15. R. S. Sutton, A. G. Barto, *Reinforcement Learning: An Introduction* (MIT Press, Massachusetts, 1998).
16. T. Tsuchiya, T. Tsuruoka, S.-J. Kim, K. Terabe, M. Aono, Ionic decision-maker created as novel, solid-state devices, *Sci. Adv.* **4**, eaau2057 (2018).
17. T. Niiyama, G. Furuhata, A. Uchida, M. Naruse, S. Sunada, Lotka-volterra competition mechanism embedded in a decision-making method, *J. Phys. Soc. Jpn.* **89**, 014801 (2020).
18. W. J. Freeman, Simulation of chaotic EEG patterns with a dynamic model of the olfactory system, *Biol. Cybern.* **56**, 139-150 (1987).
19. I. Tsuda, Toward an interpretation of dynamic neural activity in terms of chaotic dynamical systems, *Behav. Brain Sci.* **24**, 793-810 (2001).
20. I. Tsuda, Chaotic itinerancy and its roles in cognitive neurodynamics, *Curr. Opin. Neurobiol.* **31**, 67-71 (2015).
21. I. Tsuda, Chaotic itinerancy as a dynamical basis of hermeneutics in brain and mind, *World Futures* **32**, 167-184 (1991).
22. I. Tsuda, E. Koerner, H. Shimizu, Memory dynamics in asynchronous neural networks, *Prog. Theor. Phys.* **78**, 51-71 (1987).
23. M. Adachi, K. Aihara, Associative dynamics in a chaotic neural network, *Neural Netw.* **10**, 83-98 (1997).
24. Y. Kuniyoshi, S. Sangawa, Early motor development from partially ordered neural-body dynamics: experiments with a cortico-spinal-musculo-skeletal model, *Biol. Cybern.* **95**, 589-605 (2006).
25. T. Ikegami, Simulating active perception and mental imagery with embodied chaotic itinerancy," *J. Conscious. Stud.* **14**, 111-125 (2007).
26. J. Park, H. Mori, Y. Okuyama, M. Asada, Chaotic itinerancy within the coupled dynamics between a physical body and neural oscillator networks, *PLoS ONE* **12**, e0182518 (2017).
27. K. Kaneko, Clustering, coding, switching, hierarchical ordering, and control in a network of chaotic elements, *Physica D* **41**, 137-172 (1990).
28. K. Ikeda, K. Otsuka, K. Matsumoto, Maxwell-bloch turbulence, *Prog. Theor. Phys. Suppl.* **99**, 295-324 (1989).
29. T. Aida, P. Davis, Oscillation modes of laser diode pumped hybrid bistable system with large delay and application to dynamical memory, *IEEE J. Quantum Electron.* **28**, 686-699 (1992).
30. K. Inoue, K. Nakajima, Y. Kuniyoshi, Designing spontaneous behavioral switching via chaotic itinerancy, *Sci. Adv.* **6**, eabb3989 (2020).

31. T. Sano, Antimode dynamics and chaotic itinerancy in the coherence collapse of semiconductor lasers with optical feedback, *Phys. Rev. A* **50**, 2719-2726 (1994).
32. I. Fischer, G. H. M. van Tartwijk. A. M. Levine, W. Elsässer, E. Göbel, D. Lenstra, Fast pulsing and chaotic itinerancy with a drift in the coherence collapse of semiconductor lasers, *Phys. Rev. Lett.* **76**, 220-223 (1996).
33. M. C. Soriano, J, García-Ojalvo, C. R. Mirasso, I. Fischer, Complex photonics: Dynamics and applications of delay-coupled semiconductors lasers, *Rev. Mod. Phys.* **85**, 421-470 (2013).
34. W.-S. Lam, W. Ray, P. N. Guzdar, R. Roy, Measurement of Hurst exponents for semiconductor laser phase dynamics, *Phys. Rev. Lett.* **94**, 010602 (2005).
35. J. Ohtsubo, *Semiconductor Lasers: Stability, Instability and Chaos* (Springer, Ed. 4, 2017).
36. A. Uchida, *Optical Communication with Chaotic Lasers: Applications of Non-linear Dynamics and Synchronization* (Wiley-VCH, Weinheim, 2012).
37. A. Uchida, Y. Liu, I. Fischer, P. Davis, T. Aida, Chaotic antiphase dynamics and synchronization in multi-mode semiconductor lasers, *Phys. Rev. A* **64**, 023801 (2001).
38. Y. Liu, P. Davis, Adaptive mode selection based on chaotic search in Fabry-Perot laser diode, *Int. J. Bifurcation Chaos* **8**, 1685-1691 (1998).
39. V. Kovanis, A. Gavrielides, T. B. Simpson, J. M. Liu, Instabilities and chaos in optically injected semiconductor lasers, Appl. Phys. Lett. **67**, 2780–2782 (1995).
40. S. K. Hwang, J. M. Liu, Dynamical characteristics of an optically injected semiconductor laser, *Opt. Commun.* **183**, 195-205 (2000).
41. S. Wieczorek, B. Krauskopf, D. Lenstra, Mechanisms for multistability in a semiconductor laser with optical injection. *Opt. Commun.* **183**, 215-226 (2000).
42. F. Rogister, P. Mégret, O. Deparis, M. Blondel, Coexistence of in-phase and out-of-phase dynamics in a multi-mode external-cavity laser diode operating in the low-frequency fluctuations regime, *Phys. Rev. A* **62**, 061803 (2000).
43. R. Lang, K. Kobayashi, "External optical feedback effects on semiconductor injection laser properties," *IEEE J. Quantum Electron.* **16**, 347-355 (1980).
44. K. Kanno, A. Uchida, M. Bunsen, Complexity and bandwidth enhancement in unidirectionally coupled semiconductor lasers with time-delayed optical feedback, *Phys. Rev. E* **93**, 032206 (2016).
45. S.-J. Kim, M. Aono, M. Hara, Tug-of-war model for the two-bandit problem: nonlocally-correlated parallel exploration via resource conservation, *BioSystems* **101,** 29-36 (2010).
46. S.-J. Kim, M. Aono, Amoeba-inspired algorithm for cognitive medium access, *NOLTA* **5**, 198-209 (2014).
47. S.-J. Kim, M. Aono, E. Nameda, Efficient decision-making by volume-conserving physical object, *New J. Phys.* **17**, 083023 (2015).
48. S.-J. Kim, M. Naruse, M. Aono, Harnessing the computational power of fluids for optimization of collective decision making, *Philosophies* **1**, 245-260 (2016).
49. P. Auer, N. Cesa-Bianchi, P. Fischer, Finite-time analysis of the multi-armed bandit problem, *Mach. Learn.* **47**, 235-256 (2002).
50. N. Narisawa, N. Chauvet, M. Hasegawa, M. Naruse, Arm order recognition in multi-armed bandit problem with laser chaos time series, *Sci. Rep.* **11**, 4459 (2021).
51. J. D. Hart, Y. Terashima, A. Uchida, G. B. Baumgartner, T. E. Murphy, R. Roy, Recommendations and illustrations for the evaluation of photonic random number generators, *APL Photonics* **2**, 090901 (2017).
52. R. Kleinberg, Nearly tight bounds for the continuum-armed bandit problem, *Advances in Neural Information Processing Systems* **17**, 697–704 (2004).

53. S. Takeuchi, M. Hasegawa, K. Kanno, A. Uchida, N. Chauvet, M. Naruse, Dynamic channel selection in wireless communications via a multi-armed bandit algorithm using laser chaos time series, *Sci. Rep.* **10**, 1574 (2020).
54. X. Chen, B. Li, R. Proietti, H. Lu, Z. Zhu, S. J. B. Yoo, DeepRMSA: A deep reinforcement learning framework for routing, modulation and spectrum assignment in elastic optical networks, *IEEE J. Lightwave Technol.* **37**, 4155-4163 (2019).
55. K. Wada, N. Kitagawa, T. Matsuyama, The degree of temporal synchronization of the pulse oscillations from a gain-switched multimode semiconductor laser, *Materials* **10**, 950 (2017).
56. I. V. Koryukin, P. Mandel, Dynamics of semiconductor lasers with optical feedback: comparison of multi-mode models in the low-frequency fluctuation regime, *Phys. Rev*. A **70**, 053819 (2004).
57. R. Iwami, K. Kanno, A. Uchida, Chaotic mode competition dynamics in a multi-mode semiconductor laser with optical feedback and injection (in preparation).

**Acknowledgments**

**Funding:** This study was supported in part by JSPS KAKENHI (JP19H00868, JP20K15185, JP20H00233), JST CREST (JPMJCR17N2), and the Telecommunications Advancement Foundation.

**Author contributions:** All the authors contributed to the development and implementation of the concept. R. I. performed the numerical simulations and analyzed the data. R. I., T. M., K. K., S. S., M. N., and A. U. contributed to the discussion of the results. R. I., S. S., M. N., and A. U. contributed to manuscript writing.

**Competing interests:** The authors declare that they have no competing interests.

**Data and materials availability:** All data are available in the main text.

**Figures and Tables**

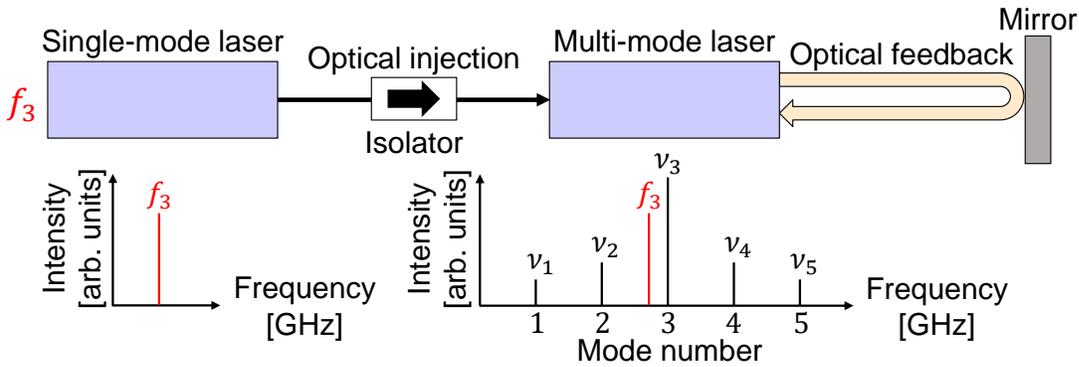

**Fig. 1. Multi-longitudinal-mode semiconductor laser with optical feedback and injection.** Multi-mode semiconductor laser is subjected to a time-delayed optical feedback by an external mirror. Optical signal from the single-mode semiconductor laser is injected into the multi-mode semiconductor laser via an optical isolator. Example of five longitudinal modes is shown.

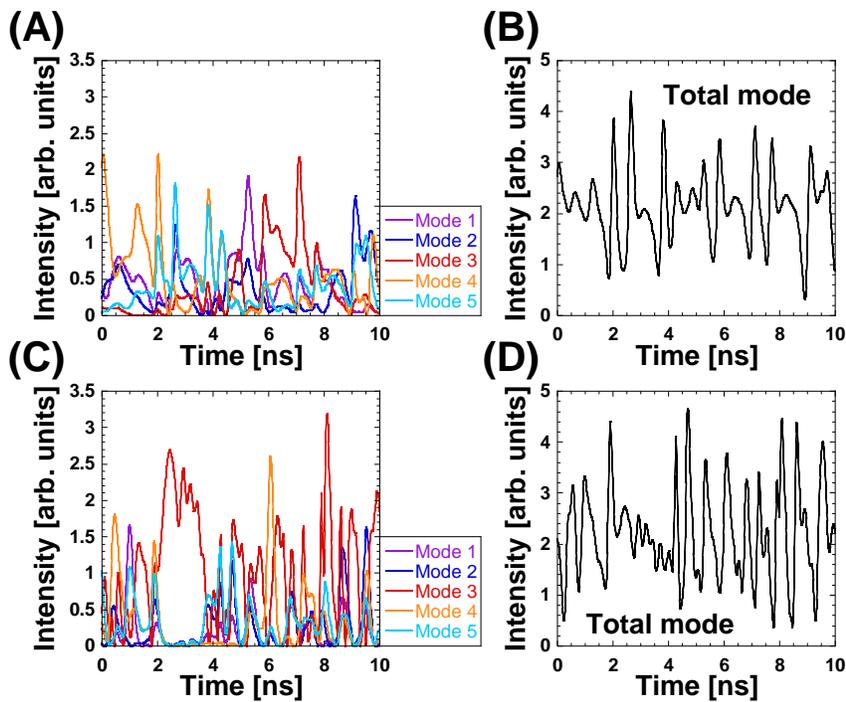

**Fig. 2. Temporal waveforms of multi-mode semiconductor laser with optical feedback.** (A), (B) No optical injection, and (C), (D) with optical injection to mode 3 (injection strength, $\kappa_{inj,3} = 6.0$ ns$^{-1}$). (A), (C) Modal intensities of the five modes and (B), (D) total intensity are shown.

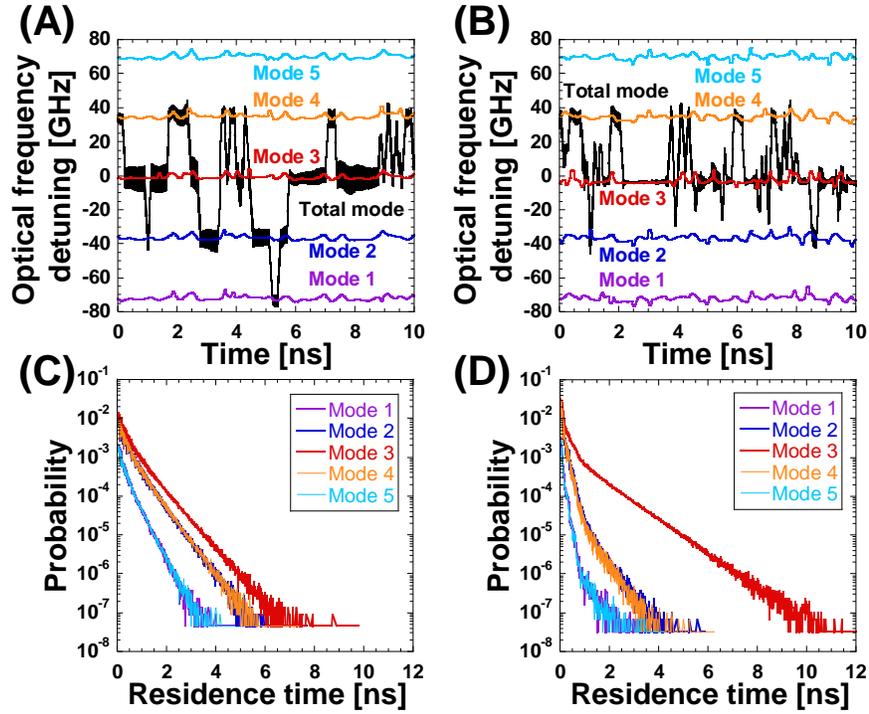

**Fig. 3. Chaotic itinerancy of total intensity among five modes with different oscillation frequencies.** (A), (B) Temporal waveforms of the total and modal intensities, and (C), (D) probabilities of the residence time of total intensity in each mode. (A), (C) No optical injection and (B), (D) with optical injection to mode 3 (the injection strength $\kappa_{inj,3} = 6.0$ ns$^{-1}$).

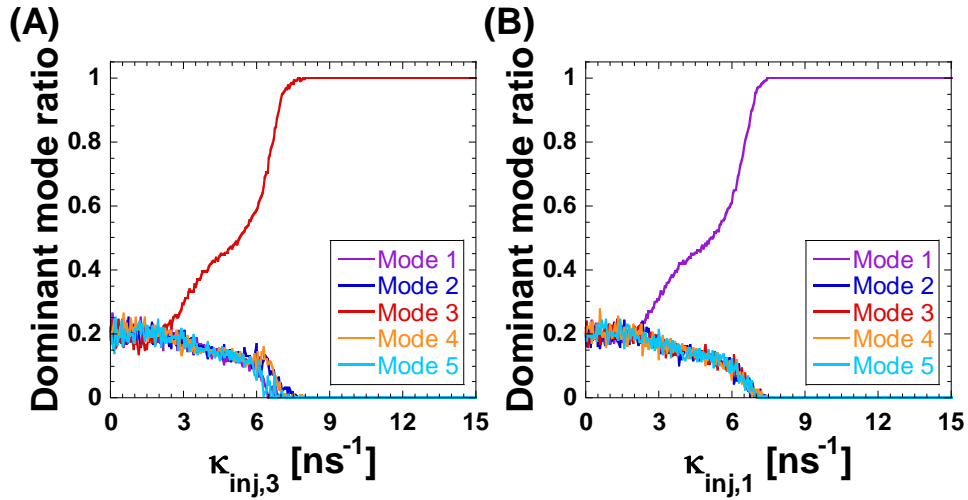

**Fig. 4. Dominant mode ratio for five modes as a function of the optical injection strength.** Optical signal is injected into (A) mode 3 and (B) mode 1. The dominant mode ratio is calculated from the temporal waveforms over 20 μs for each optical injection strength.

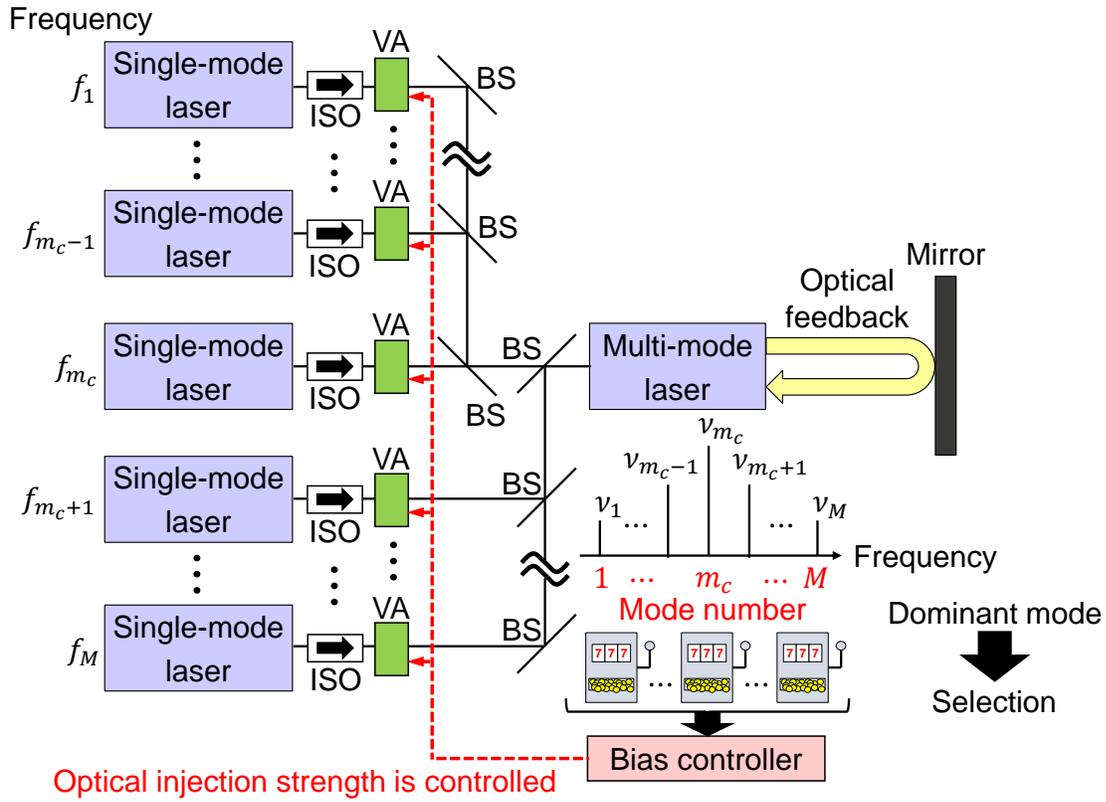

**Fig. 5. Schematic of decision-making method using multi-mode semiconductor laser with optical feedback and injection.** $v_m$: frequency of the $m$-th longitudinal mode of the multi-mode semiconductor laser, $f_m$: frequency of the single-mode semiconductor laser for optical injection to the $m$-th mode, BS: beam splitter, ISO: optical isolator, VA: variable attenuator.

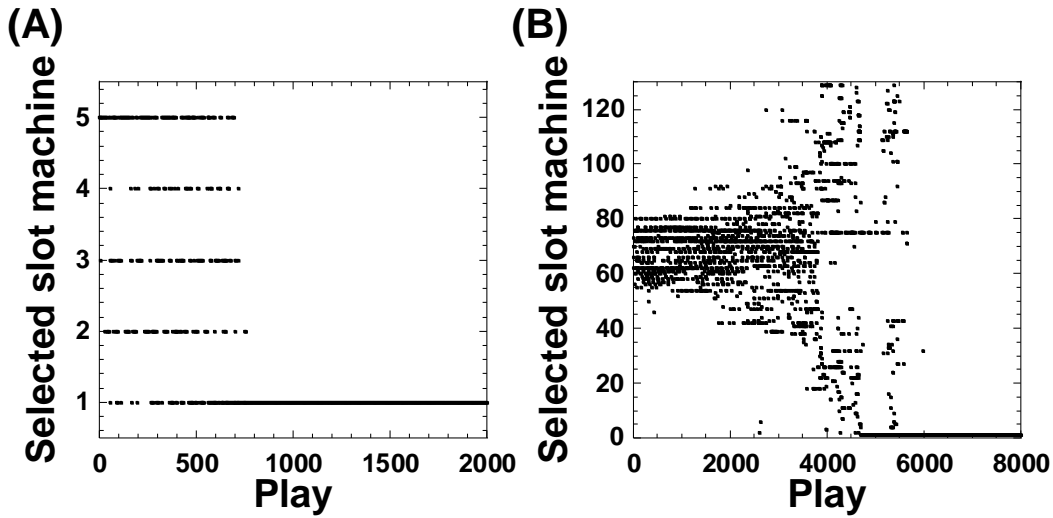

**Fig. 6. Result of slot machine selection as a function of the number of plays.** The numbers of slot machines are (A) $M = 5$ and (B) $M = 129$. The hit probability of the slot machine 1 is set to 0.9 and those of the other slot machines are set to 0.7.

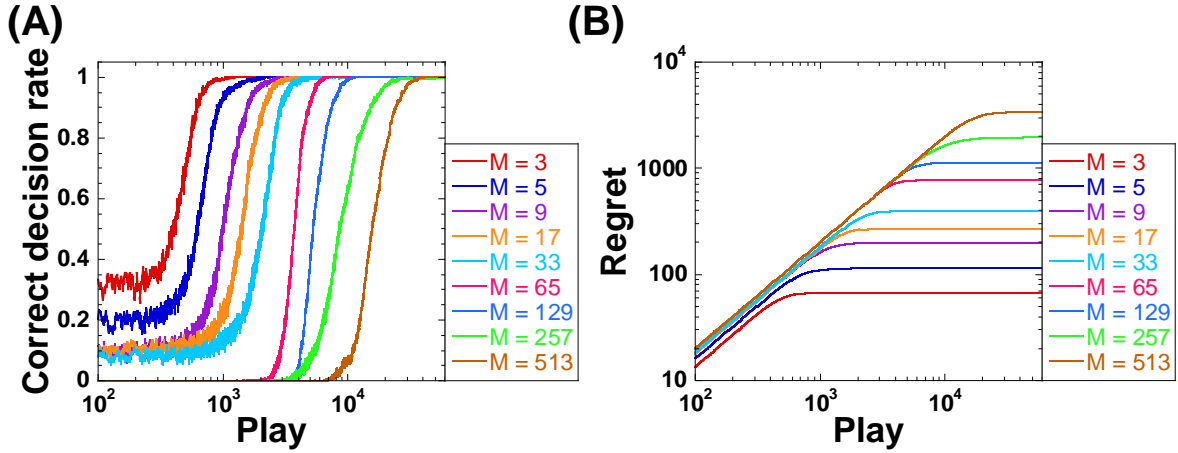

**Fig. 7. Decision-making performance for different numbers of slot machines from $M = 3$ to $M = 513$ ($M = 2^n + 1$).** (A) Correct decision rate (CDR) and (B) regret as functions of the number of plays. The hit probability of slot machine 1 is set to 0.9 and those of the other slot machines are set to 0.7. Logarithmic scale is used (A) in the horizontal axis and (B) in the vertical and horizontal axes.

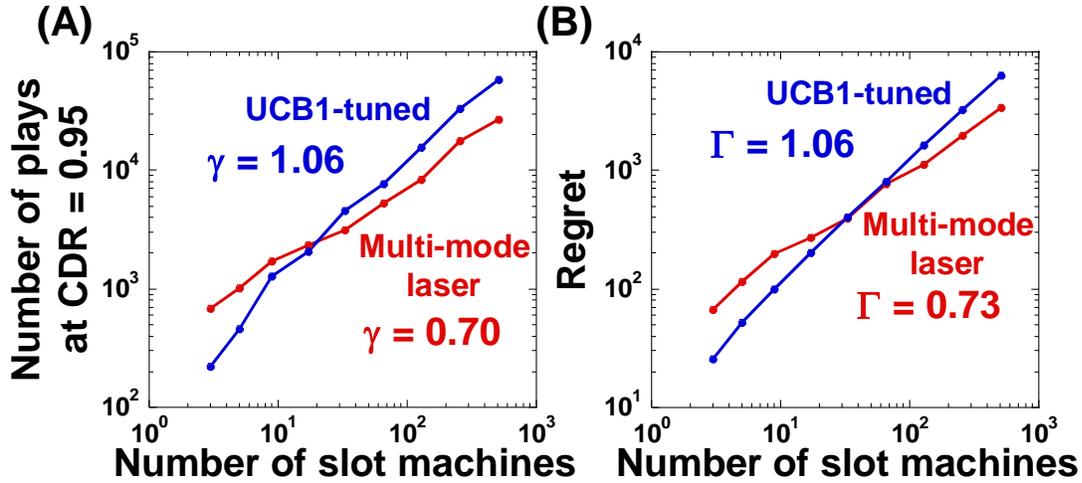

**Fig. 8. Comparison of scalability of the multi-mode semiconductor laser (red) and UCB1-tuned software algorithm (blue).** The hit probability of slot machine 1 is set to 0.9 and those of the other slot machines are set to 0.7. (A) Scalability between the number of plays required for the CDR of 0.95 and the number of slot machines. The scaling exponents are 0.70 and 1.06 for the multi-mode laser and UCB1-tuned method, respectively. (B) Scalability between the regret at the 60000-th play and the number of slot machines. The scaling exponents are 0.73 and 1.06 for the multi-mode laser and UCB1-tuned method, respectively.

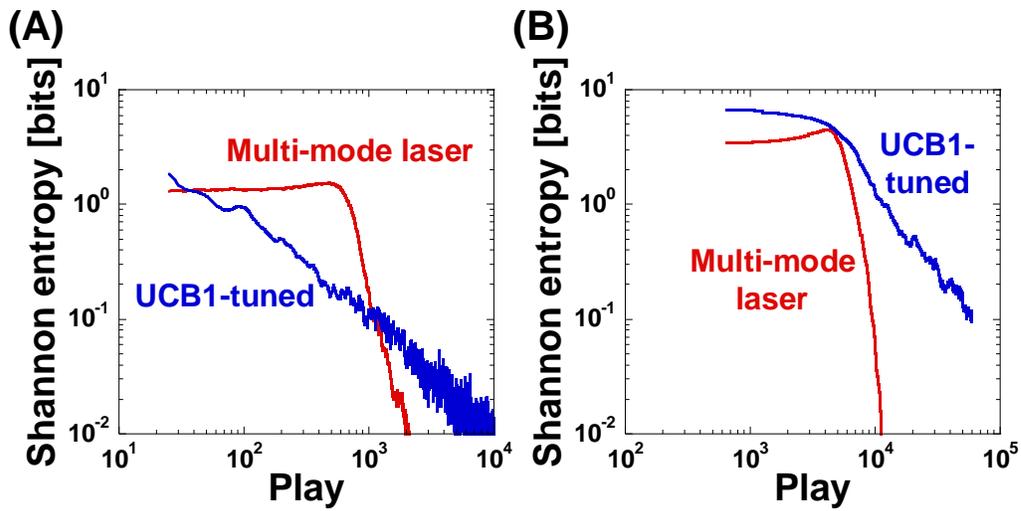

**Fig. 9. Shannon entropy of probabilities of slot machine selection as a function of number of plays for the multi-mode laser (red) and UCB1-tuned software algorithm (blue).** The numbers of slot machines are (A) $M = 5$ and (B) $M = 129$. The entropy of the $p$-th play is calculated from the selection probability using the last $5M$ plays of the $p$-th play and averaged over 500 cycles.

**Table 1. Parameter values used in the numerical simulations of the multi-mode semiconductor laser with optical feedback and injection.**

| Symbol | Parameter | Value |
| --- | --- | --- |
| $G_N$ | Gain coefficient at central mode $m_c$ | $8.40 \times 10^{-13}$ m$^3$s$^{-1}$ |
| $N_0$ | Carrier density at transparency | $1.40 \times 10^{24}$ m$^{-3}$ |
| $\alpha$ | Linewidth enhancement factor | 3.0 |
| $\varepsilon$ | Gain saturation coefficient | $2.5 \times 10^{-23}$ |
| $\tau_p$ | Photon lifetime | $1.927 \times 10^{-12}$ s |
| $\tau_s$ | Carrier lifetime | $2.04 \times 10^{-9}$ s |
| $\kappa$ | Optical feedback strength of multi-mode semiconductor laser | $4.411 \times 10^{9}$ s$^{-1}$ |
| $\kappa_{inj,m}$ | Optical injection strength from single-mode semiconductor laser to mode $m$ | Variable ($0.0 \sim 15.0 \times 10^{9}$ s$^{-1}$) |
| $\tau$ | Round-trip time of light in external cavity of multi-mode semiconductor laser | $1.001 \times 10^{-8}$ s |
| $J$ | Injection current | $1.11\ J_{th}$ |
| $\Delta\nu$ | Frequency of longitudinal mode spacing | $3.55 \times 10^{10}$ Hz |
| $\nu_{mc}$ | Frequency of central mode $m_c$ | $1.951 \times 10^{14}$ Hz |
| $\Delta\nu_g$ | Frequency width of gain profile | $1.419 \times 10^{14}$ Hz |
| $\Delta f_{mc}$ | Initial optical frequency detuning between single-mode laser and central mode $m_c$ in multi-mode laser | $-4.0 \times 10^{9}$ Hz |
| $A_s$ | Steady-state solution of electric-field amplitude of single-mode semiconductor laser | $1.438 \times 10^{10}$ |
| $J_{th} = N_{th} / \tau_s$ | Injection current at lasing threshold | $9.891 \times 10^{32}$ m$^{-3}$ s$^{-1}$ |
| $N_{th} = N_0 + 1 / (G_N \tau_p)$ | Carrier density at lasing threshold | $2.018 \times 10^{24}$ m$^{-3}$ |